\documentclass[a4paper]{article}

\usepackage{INTERSPEECH2018}

\usepackage{subcaption}

\usepackage{array}
\usepackage{booktabs}
\usepackage{amsmath}
\usepackage{url}
\usepackage{todonotes}
\usepackage{tablefootnote}

\title{Unspeech: Unsupervised Speech Context Embeddings}
\name{Benjamin Milde, Chris Biemann}
\address{Language Technology, Universit\"at Hamburg}
\email{\{milde,biemann\}@informatik.uni-hamburg.de}

\begin{document}

\maketitle
\begin{abstract}

We introduce \textit{"Unspeech"} embeddings, which are based on unsupervised learning of context feature representations for spoken language. The embeddings were trained on up to 9500 hours of crawled English speech data without transcriptions or speaker information, by using a straightforward learning objective based on context and non-context discrimination with negative sampling. We use a Siamese convolutional neural network architecture to train Unspeech embeddings and evaluate them on speaker comparison, utterance clustering and as a context feature in TDNN-HMM acoustic models trained on TED-LIUM, comparing it to i-vector baselines. Particularly decoding out-of-domain speech data from the recently released Common Voice corpus shows consistent WER reductions. We release our source code and pre-trained Unspeech models under a permissive open source license. 
\end{abstract}

\section{Introduction}

Variance and variability in recordings of speech and its representations are a common problem in automatic speech processing tasks. E.g. speaker, environment characteristics and the type of microphone will cause large differences in typical speech representations (e.g. FBANK, MFCC), making direct similarity comparisons difficult. We can describe such factors of variance also as the context of an utterance; speech sounds that occur close in time share similar contexts. Based on this idea, we propose to learn representations of such contexts in an unsupervised way, without needing further speaker IDs, channel information or transcriptions of the data.

Recent acoustic models for automatic speech recognition (ASR) incorporate some form of (deep) neural network that can learn to deal with part of this variance by using supervised training data in combination with the ability to learn representations as part of the model. A growing trend is to incorporate larger context views of the data explicitly into the neural network. In Deep Neural Network Hidden Markov Model (DNN-HMM) hybrids, fixed-length speaker embeddings like i-vectors are made available for the neural network as additional input features \cite{saon2013speaker}. Typically, larger temporal windows than single speech frames are also used as input to the neural network to make context available to local predictions. This can e.g. be achieved by either stacking consecutive speech frames or by using Time-Delayed Neural Networks (TDNNs) ~\cite{waibel1990phoneme, peddinti2015time}. %

On the other hand, \textit{"Unspeech"} embedding models embed a window of speech into a fixed-length vector so that corresponding points are close, if they share similar contexts. Unsupervised training of the embedding function is inspired by negative sampling in word2vec \cite{mikolov2013distributed} -- where words that share a similar meaning are embedded in similar regions in a dense vector space. In this work, we demonstrate that the learned Unspeech context embeddings encode speaker characteristics and also can be used to cluster a speech corpus. As an additional context input feature, they can also improve supervised speech recognition tasks with TDNN-HMM acoustic models, in particular when adaptation to out-of-domain data is needed. %

\section{Related Work}

Speaker embeddings and phonetic embeddings are two major groups of proposed embeddings in speech: While speaker embeddings seek to model utterances from the same speaker so that they share similar regions in a dense vector space, in phonetic embeddings, the same or similar phonetic content is close.

I-vectors \cite{dehak2011front} are well-known, popular speaker vectors. Recently, supervised neural network-based speaker embeddings also succeeded to show good speaker discriminative properties \cite{snyder2016deep, snyder2017deep, li2017deep}, particularly on short utterances. Bengio and Heigold proposed supervised word embeddings for speech recognition \cite{bengio2014word}, where words are nearby in the vector space if they sound alike. Kamper et al. \cite{kamper2015unsupervised} showed that auto-encoders can also be used in conjugntion with top-down information for  unsupervised phonetic representation learning in speech. Chung et al. \cite{chung2016audio} proposed audio word2vec, based on sequence-to-sequence auto-encoders trained on a dictionary of isolated spoken words. By analogy of auto-encoders, Pathak et al. introduced context encoders \cite{pathak2016context}, a class of models that learn context embeddings in images. There is also growing interest in representation learning on non-speech audio by using learning objectives directly related to contexts. Jansen et al. \cite{jansen2017unsupervised} encoded the notion that (non-speech) sounds occurring in context are more related.

Bromley et al. \cite{bromley1994signature} introduced Siamese neural networks: two (time-delayed) neural networks that embed digital signatures and a learning objective based on discriminating between true and false signatures. This idea has recently been revisited in the context of joint phoneme and speaker embeddigs learning in a weakly supervised setting, where speaker annotation, same word information and segmentation is available \cite{synnaeve2014weakly, zeghidour2016joint}. Gutmann et al. \cite{gutmann2010noise} introduced Noise-Contrastive Estimation (NCE), an estimator based on discriminating between observed data and some artificially generated noise. Jati et al.  proposed Speaker2vec \cite{jati2017speaker2vec} for speaker segmentation, with unsupervised training using a neural encoder/decoder. Very recently and in parallel to our efforts, Jati et al. also proposed (unsupervised) neural predictive coding to learn speaker characteristics \cite{jati2018neural}.

In \cite{jin1997automatic} unsupervised speaker clustering was proposed to yield labels for speaker adaptation in acoustic models, based on the idea that consecutive windows of speech are likely from the same speaker. Several forms of context/speaker embeddings have also been used for (speaker) adaptation in state-of-the-art speech recognition acoustic models: i-vector speaker embeddings are by far the most popular \cite{saon2013speaker, senior2014improving, miao2015speaker}. Vesely et al. proposed sequence summary neural networks for speaker adaptation, where utterance context vectors are averaged from the speech feature representation \cite{vesely2016sequence}. Gupta et al. \cite{gupta2017visual} showed that visual features, in the form of activations from a pre-trained ConvNet for object detection on videos can also be used as context vectors in the acoustic model.

\section{Proposed Models}

\begin{figure}[t!]
    \centering
        \centering
        \includegraphics[width=0.9\linewidth]{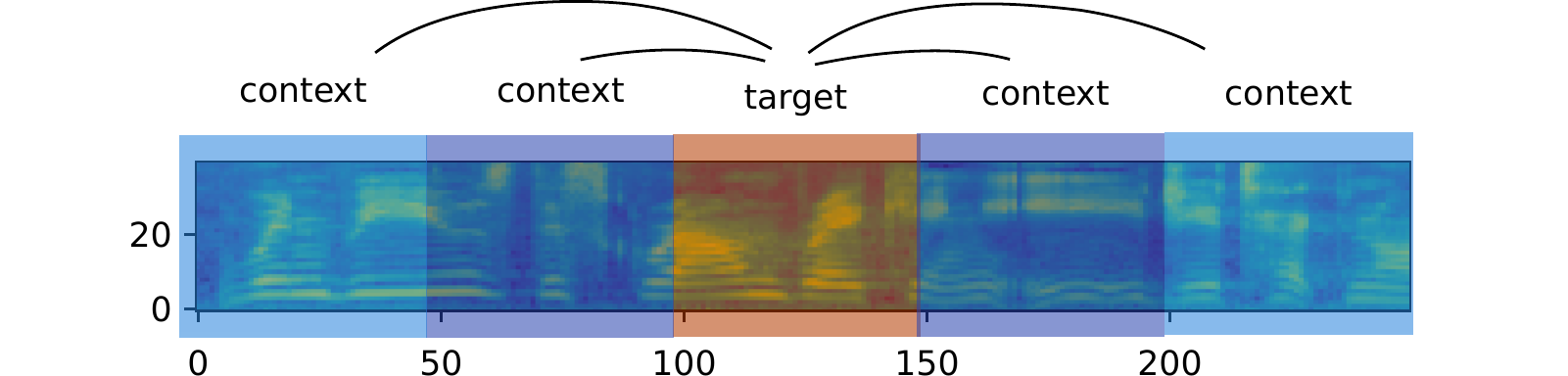}
        \caption{The initial sequence with unnormalized FBANK vectors: we choose one target window and two left and right contexts. All windows are of the same size. }
            \label{fig:unspeech_sampling}
\end{figure}

\begin{figure}[t!]
        \centering
        \includegraphics[width=1\linewidth]{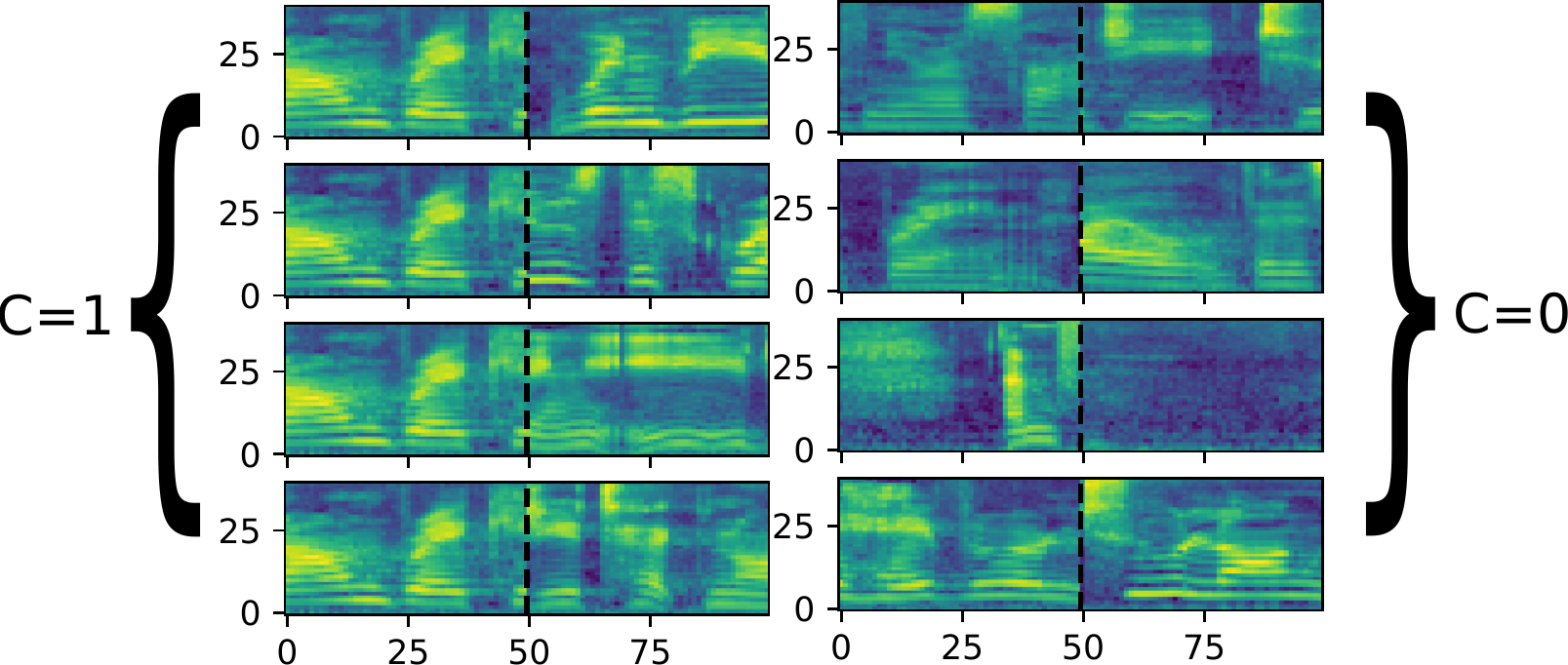}
    \caption{Sampling examples for two left contexts and two right contexts from the figure above. Positive example pairs are of class $C = 1$, negative sampled pairs are $C = 0$. In this example, each window has a size of 50 FBANK frames (0.5 seconds).}
        \label{fig:unspeech_left_right_labels}
    \vspace{-1mm}
\end{figure}

We construct an artificial binary classification task with logistic regression, where two fixed sized windows are compared. One target window can have multiple context windows, depending on the number of left and right contexts. For every left and right context, a pair with the target is created. Figure \ref{fig:unspeech_sampling} illustrates this with two right and left contexts, yielding four positive contexts and four randomly sampled negative contexts. For the target window we denote $emb_t$ as the target embedding transformation, taking a window of FBANK features and producing a fixed sized output vector, $emb_c$ as the embedding of a true context and $emb_{neg}$ as the embedding of a randomly sampled context. The pair of (embedded) speech windows is considered to be of class $C = 1$ if one window is the context of the other, or $C = 0$ if they are not. For $C = 1$, we sample the pairs from consecutive windows, for $C=0$ we use negative sampling to construct a pair of speech windows that are unrelated with high probability: We uniformly sample a random utterance $u$ and then uniformly a random position in $u$.

\subsection{Objective Function}

With the scalar x as the output of the model for a particular data point, $\sigma$ the sigmoid function and C its true class $\in (0,1)$, the logistic loss for a binary classification task is: 

\vspace{-2mm}

\begin{equation}
loss(x,C) = C(-log(\sigma(x))) + (1 - C)(-log(1 - \sigma(x)))
\end{equation}

with $x=emb_t^T emb_c$, the dot product over target and context embedding transformations if $C=1$ and $x={emb_{neg1}}_i^T {emb_{neg2}}_i$, the dot product over two negative sampled embedding transformations if $C=0$, for $k=$ number of negative samples we can thus obtain:

\begin{equation}
\begin{split}
NEG_{loss} = -k \cdot log(\sigma(emb_t^T emb_c)) \\
- \sum_{i=1}^{k} log(1 - \sigma(emb_{{neg1}_i}^T emb_{{neg2}_i})) 
\end{split}
\end{equation}

Note that in the similar NCE loss formulation \cite{gutmann2010noise}, $P(C=1) = P(C=0) = \frac{1}{2}$, i.e. the number of data points where $C=1$ and $C=0$ are the same, while we could have more negative than positive target/context embedding pairs, depending on $k$. Instead, for $C=1$ we multiply with $k=$ the number of negative samples, to penalize errors on positive and negative context pairs equally. Another difference is that we sample two unrelated embedding windows instead of one.

\subsection{Model Architecture}

\begin{figure}[th!]
    \centering
    \includegraphics[width=0.4\textwidth]{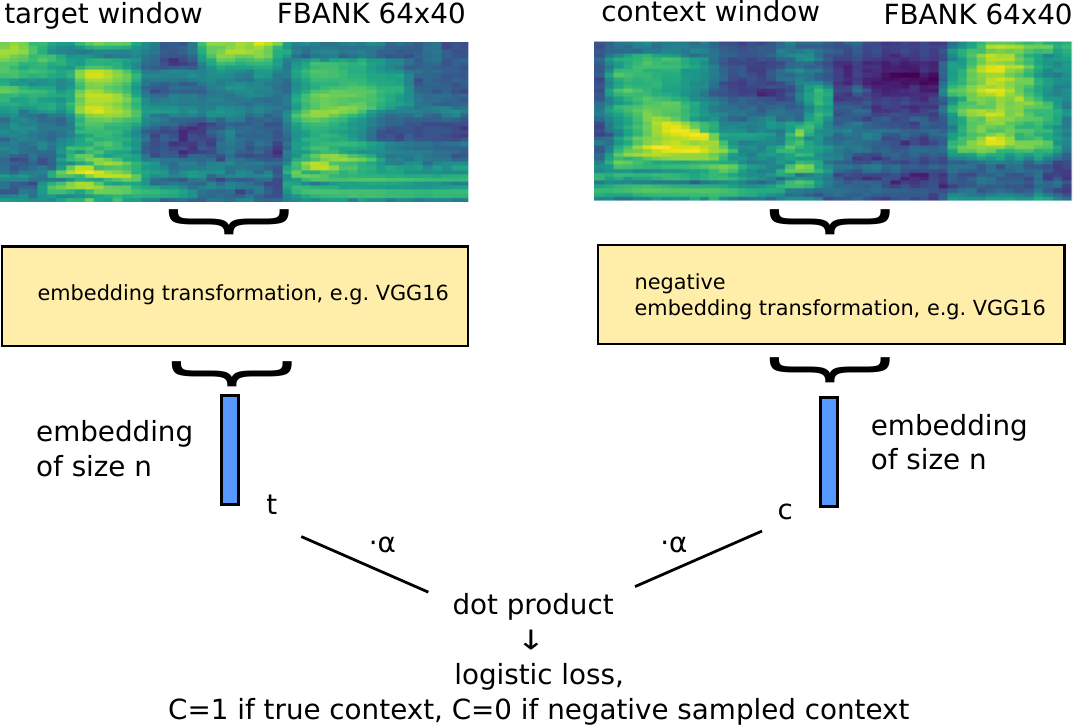}
    \caption{Unspeech embeddings are trained using a Siamese neural network architecture combined with a dot product. We use the VGG16A network as embedding transformation in the yellow boxes (a convolutional neural network with 16 layers).} %
    \label{fig:unspeech_neg_model}
    \vspace{-1mm}
\end{figure}

Figure \ref{fig:unspeech_neg_model} shows our architecture for FBANK input features. We project two dimensional input windows into two fixed sized embeddings, which are combined with a dot product. Since this is sensitive to the scaling of output embeddings, we multiply them with a single scalar parameter $\alpha$, which is trained with the rest of the network. Direct normalization to unit length (making the dot product equivalent to a cosine distance) did hamper convergence of the loss and was discarded early on.

Many different architectures are possible for converting the input representation into a fixed sized embedding, but we mainly evaluated with a VGG-style ConvNet architecture (Model A in \cite{simonyan2014very}), as it is well established and it can exploit the two dimensional structure in the FBANK signal. We share the weights of the convolutional layers of both embedding transformations, but keep the fully connected layers separate. Dropout is used for fully connected layers (0.1) and L2 regularization is used on the weights (0.0001), for all experiments we optimize with ADAM \cite{DBLP:journals/corr/KingmaB14}. We make use of leaky ReLUs \cite{maas2013rectifier}.

\section{Evaluation}

\vspace{-2mm}

\begin{table}[h]
\centering
\caption{Comparison of English speech data sets.}
\label{tab:datasets}
\scalebox{0.85}{
\begin{tabular}{l|lll|lll}
 & \multicolumn{3}{c|}{\textbf{hours}} 
 &  \multicolumn{3}{c}{\textbf{speakers}}    \\ \hline
\textbf{dataset} & \textbf{train} & \textbf{dev} & \textbf{test} & \textbf{train} & \textbf{dev} & \textbf{test} \\ \hline
TED-LIUM V2 & 211 & 2 & 1 & 1273+3 & 14+4 & 13+2 \\ 
Common Voice V1 & 242 & 5 & 5 & 16677 & 2728 & 2768  \\ 
TEDx (crawled)& \multicolumn{3}{l|}{9505} & \multicolumn{3}{l}{41520 talks}\\
\end{tabular}
}
\vspace{-2mm}
\end{table}

Table \ref{tab:datasets} characterizes the datasets we used in our evaluation. TED-LIUM V2 \cite{rousseau2014enhancing} has a comparatively small number of speakers, especially in the development and test set of the corpus. TED-LIUM and Common Voice \cite{mozilla2018} are segmented at the utterance level, both are similar in the number of hours. In Common Voice, volunteers from all over the world recorded predefined prompts, in TED-LIUM utterances are segmented from and aligned to TED talks. In order to explore large-scale training, we also downloaded all TEDx talks from 01-01-2016 until 03-01-2018 from YouTube, giving us 41520 talks (0.5 TB compressed audio data) with a total of 9505 hours of unannotated audio. While the majority of TEDx talks are in English, a very small number of them are in other languages or contain only music. We did not segment or clean the TEDx data.

\subsection{Same/different Speaker Experiment}

In the same/different speaker experiment, we evaluate a binary classification task: given two utterances, are they uttered from the same or different speakers? Our hypothesis is that Unspeech embeddings can be used for this task, because one strategy to discriminate samples of true contexts from negative sampled ones is modelling speaker traits. In Table \ref{tab:samedifferent} we show equal error rates (EER)\footnote{The error rate where the number of false positive and false negatives is the same, calculated using pyannote.metrics \cite{Bredin2017a}} on same/different speaker comparisons of all utterance pairs, limiting the number of speakers to 100 in the train sets of TED-LIUM and Common Voice corpus in this experiment. The embedding dimension is 100 in all experiments, we train Unspeech models with different target window widths (32, 64, 128) and i-vectors are trained/extracted with Kaldi \cite{povey2011kaldi}. For all experiments, we use two left and two right context windows.

\begin{table}[h]
\centering
\caption{Equal error rates (EER) on TED-LIUM V2 -- Unspeech embeddings correlate with speaker embeddings. }
\label{tab:samedifferent}
\begin{tabular}{l|l|l|l}

\textbf{Embedding} & \multicolumn{3}{c}{\textbf{EER}}  \\ \hline
\raggedright{\textbf{TED-LIUM:}} & \textbf{train}  & \textbf{dev} & \textbf{test} \\ \hline
(1) i-vector & 7.59\% & \textbf{0.46\%} & 1.09\% \\ 
(2) i-vector-sp &  \textbf{7.57\%} & 0.47\% & \textbf{0.93\%}  \\
(3) unspeech-32-sp & 13.84\% & 5.56\% & 3.73\% \\
(4) unspeech-64 & 15.42\% & 5.35\% & 2.40\%  \\
(5) unspeech-64-sp & 13.92\%  & 3.4\% & 3.31\%\\
(6) unspeech-64-tedx & 19.56\%  & 7.96\%  & 4.96\% \\
(7) unspeech-128-tedx & 20.32\%  & 5.56\%  & 5.45\% \\
\end{tabular}
\vspace{-2mm}
\end{table}

The distance function $d_1(a,b)=\sigma({emb_t}(a)^{T}{emb_c}(b))$ to compare two segments $a$ and $b$ correspond to the distance function in the Unspeech training process. The cosine distance, or equally after normalization to unit length, the Euclidean distance on vectors produced by $emb_t$ also produces good comparison results, so that $d_2(a,b) = ||{emb_t}(a)-{emb_t}(b)||$ can be used for comparing two Unspeech segments. Sequences that are longer than the trained target window can be windowed and averaged to obtain a single vector for the whole sequence, since vectors that are close in time share contexts and correlate highly.
However, EER on i-vectors trained with supervised speaker labels compared with the cosine distance (results with $d_2(a,b)$ are identical after normalization) are lower than on Unspeech embeddings with $d_2(a,b)$ (1,2 vs 3,4,5). Training Unspeech on TEDx talks instead of TED-LIUM does also produce higher EER as a speaker embedding (6,7). "-sp" denotes training on speed-perturbed data: adding copies of the raw training data at 0.9 and 1.1 playing speed, as recommended in \cite{ko2015audio}. %

\subsection{Clustering Utterances} \label{sec:clusterunspeech}

\begin{table*}[t]
\centering
\caption{Comparing clustered utterances from TED-LIUM using i-vectors and (normalized) Unspeech embeddings with speaker labels from the corpus. "-sp" denotes embeddings trained with speed-perturbed training data.}
\label{tab:clustering}
\begin{tabular}{l|l|l|l|l|l|l|l|l|l|l|l|l}
\textbf{Embedding} & \multicolumn{3}{c|}{\textbf{Num. clusters}} & \multicolumn{3}{c|}{\textbf{Outliers}} & \multicolumn{3}{c|}{\textbf{ARI}} & \multicolumn{3}{c}{\textbf{NMI}} \\ %
\hline
& \textbf{train} & \textbf{dev} & \textbf{test} & \textbf{train} & \textbf{dev} & \textbf{test} & \textbf{train} & \textbf{dev} & \textbf{test} & \textbf{train} & \textbf{dev} & \textbf{test} \\
\hline
TED-LIUM IDs & 1273 (1492) & 14 & 13 & 3 & 4 & 2 & 1.0 & 1.0 & 1.0 & 1.0 & 1.0 & 1.0 \\

i-vector &  1630 & 12 & 10 & 8699 & 1 & 2 & 0.8713 & \textbf{0.9717} & \textbf{0.9792} & 0.9605 & \textbf{0.9804} & \textbf{0.9598} \\

i-vector-sp &  1623 & 12 & 10 & 9068 & 1 & 2 & 0.8641 & \textbf{0.9717} & \textbf{0.9792} & 0.9592 & \textbf{0.9804} & \textbf{0.9598} \\

unspeech-32-sp & 1686 & 16 & 12 & 3235 & 22 & 32 & \textbf{0.9313} & 0.9456 & 0.9178 & \textbf{0.9780} & 0.9536 & 0.9146 \\

unspeech-64 &  1690 & 16 & 11 & 5690 & 14 & 21 & 0.8130 & 0.9537 & 0.9458 & 0.9636 & 0.9636 & 0.9493 \\

unspeech-64-sp &  1702 & 15 & 11 & 3705 & 23 & 25 & 0.9205 & 0.9517 & 0.9340 & 0.9730 & 0.9633 & 0.9366 \\
  
\end{tabular}

\vspace{-2.5mm}
\end{table*}

We can also use the generated vectors to cluster a corpus of utterances to gain insight into what kind of utterances get clustered together. We use HDBSCAN \cite{mcinnes2017hdbscan}, a modern hierarchical density-based clustering algorithm for our experiments since it scales very well to a large number of utterances and the number of clusters does not need to be known apriori. It uses an approximate nearest neighbor search if the comparison metric is Euclidean, making it significantly faster on a large number of utterances as compared to other speaker clustering methods that require distance computations of all utterance pairs (including greedy hierarchical clustering with BIC \cite{zhou2000unsupervised}).
We use Adjusted Rand Index (ARI) \cite{hubert1985comparing} and Normalized Mutual Information (NMI) \cite{strehl2002relationship} to compare the clusters to the speaker IDs provided by the TED-LIUM corpus in Table \ref{tab:clustering}. %
We found that Unspeech embeddings and i-vectors give a sensible number of clusters, without much tweaking of HDBSCAN's two parameters (min. cluster size, min. samples)\footnote{We use 5/3 for all experiments shown in Table \ref{tab:clustering}, but other parameters in the range 3-10 will give similar results.}. 
On the train set, Unspeech embeddings will provide slightly higher cluster scores, while i-vectors provide better scores for dev and test (that have a significantly smaller number of speakers). Unspeech-64 is slightly better than Unspeech-32 on the dev and test set. ARI is sensitive to the absolute number of outliers -- we found NMI to be a better metric to compare the results on the train set.

Taking a closer look at the clustered Unspeech embeddings, we observed that different speakers in the same talk tend to get clustered into distinct clusters (making the clustered output very often more accurate than the train speaker IDs provided in TED-LIUM), while the same speaker across different talks and also the same speaker in one talk with significantly different background noises tends to be clustered into distinct clusters. This implies that Unspeech embeds more than just speaker traits.

\subsection{Acoustic Models With Unspeech Cluster IDs}

\begin{table}[]
\centering
\caption{Comparing the effect of two speaker division baselines (One speaker per talk, one speaker per utterance) and clustering with Unspeech on WER with GMM-HMM and TDNN-HMM chain acoustic models trained on TED-LIUM.}
\label{tab:baselines_TDNN_Tedlium}
\begin{tabular}{l|c|cc|cc}
\textbf{Acoustic model} & \textbf{Spk. div.}    & \multicolumn{2}{c}{\textbf{Dev WER}} & \multicolumn{2}{c}{\textbf{Test WER}} \\
\hline
 & & \footnotesize{plain}  & \footnotesize{resc.} & \footnotesize{plain} &\footnotesize{resc.} \\
 \hline
\footnotesize{GMM-HMM}      &   per talk      &  19.2 & 18.2        &  17.6 & 16.7        \\
\footnotesize{TDNN-HMM} &         &   8.6 & 7.8      &    8.8 & 8.2      \\
\hline
\footnotesize{GMM-HMM }       &   per utt.  &  19.6 & 18.7       &   20.1 & 19.2       \\
\footnotesize{TDNN-HMM} &      & 8.5 & 7.9       &   9.3 & 9.0       \\
\hline
\footnotesize{GMM-HMM}      &   Unspeech & 18.4 & 17.4        &  17.5 & 16.5 \\
\footnotesize{TDNN-HMM} &    64  &  8.6 & 7.8        &     \textbf{8.5} & \textbf{8.1}    \\
\hline
\footnotesize{GMM-HMM}     &  Unspeech   &  18.4   &  17.5  & 17.2 & 16.4  \\
\footnotesize{TDNN-HMM} &   64-sp   &  \textbf{8.3} & \textbf{7.5}   &    8.6 & 8.2    \\          
\end{tabular}
\end{table}

We can also train acoustic models with the cluster IDs provided and use them in lieu of speaker IDs for HMM-GMM speaker adaptation and online i-vector training for the TDNN-HMM model. We use the TED-LIUM TDNN-HMM chain recipe (s5\_r2) in Kaldi \cite{povey2016purely} and show WER before (plain) and after rescoring with the standard 4-gram Cantab TED-LIUM LM (resc.). Table \ref{tab:baselines_TDNN_Tedlium} shows WER on different speaker separation strategies on the train set, with one speaker per talk being the default in the s5\_r2 recipe. All models pre-trained online i-vectors based on the given speaker IDs and use those as additional input features. The standard recipe computes a fixed affine transform on the combined input features (40 dim hi-res MFCC + 100 dim i-vector), c.f. Appendix C.6 of \cite{povey2014parallel}. GMM-HMM bootstrap models will perform about 15\% worse and TDNN-HMM trained on them will perform about 10\% worse if no speaker information is available. Using the cluster IDs from clustering Unspeech embeddings of all utterances, the baseline WER can not only be recovered, but even slightly improved upon. For all TDNN-HMM models we set the width of a layer to 1024.

\begin{table}[h!]
\centering
\caption{WER for TDNN-HMM chain models trained with Unspeech embeddings on TED-LIUM.}
\label{tab:context_emb_tedlium}
\begin{tabular}{l|ll|ll}
 \textbf{Context vector}    &  \multicolumn{2}{c|}{\textbf{Dev WER}} &  \multicolumn{2}{c}{\textbf{Test WER}} \\
 \hline
 & \footnotesize{plain}  & \footnotesize{resc.} & \footnotesize{plain} &\footnotesize{resc.} \\
\hline
(1) none      &    9.1 & 8.5     & 9.5 & 9.1         \\
(2) i-vector-sp-ted      &  \textbf{8.3} & \textbf{7.5} &   8.6 & 8.2   \\
(3) unspeech-64-sp-ted        & 9.1& 8.3  & 9.6 & 9.0      \\      
(4) unspeech-64-sp-cv			& 9.1 & 8.3	&	9.5 & 9.1	\\
(5)  unspeech-64-sp-cv + (2) &	8.4 & 7.6	&	\textbf{8.5} & \textbf{8.1}	\\
(6) unspeech-64-tedx       &   9.0 & 8.2      &   9.4 & 8.7       \\      
(7) unspeech-128-tedx     &  8.9 & 8.2      &  9.4 & 8.9        \\ 
\end{tabular}
\vspace{-4mm}
\end{table}

\subsection{Unspeech Context Vectors in TDNN-HMM models}

We can also replace the i-vector representation used in training the TDNN-HMM with the Unspeech context vector. In Table \ref{tab:context_emb_tedlium}, we selected the strongest baseline from Table \ref{tab:baselines_TDNN_Tedlium} according to the dev set (Unspeech 64-sp clusters) and show WERs on the TED-LIUM dev and test for different Unspeech context embeddings. We trained Unspeech models with different window sizes (64,128) on TED-LIUM (ted) and Common Voice V1 (cv) and computed them for every 10 frames, like the online i-vector baseline. While Unspeech embeddings can slightly improve a baseline model trained without any context vectors, with best results obtained when training on the 9500 hours of TEDx data (6,7), using i-vectors (2) yields better WERs compared to Unspeech embeddings. Combining Unspeech embeddings trained on Common Voice and i-vectors in the input representation can yield slightly lower WERs than i-vectors alone (5).

\begin{table}[]
\centering
\caption{Decoding Common Voice V1 utterances. Mozilla's open source dataset provides a challenging test set, which is out-of-domain for an acoustic model trained on TED-LIUM.}
\label{tab:common_voice_adaption}
\begin{tabular}{l|ll|ll}
 \textbf{Context vector}  & \multicolumn{2}{c|}{\textbf{Dev WER}} & \multicolumn{2}{c}{\textbf{Test WER}} \\
\hline
 & \footnotesize{plain}  & \footnotesize{resc.} & \footnotesize{plain} &\footnotesize{resc.} \\
 \hline

(1) none  &  31.2 &  29.6  &  29.9 &  28.5     \\
(2) i-vector-sp-ted  & 30.3 & 29.0  & 29.9 & 28.2   \\
(3) unspeech-64-sp-cv   &  \textbf{29.5} & \textbf{27.9}  &  \textbf{28.3} & \textbf{26.9} \\
(4) unspeech-64-sp-cv  + (2)  &  29.6 & 28.2  & 28.9 & 27.4  \\   
(5) unspeech-64-tedx &  30.2 & 28.8  &  29.2 & 27.5   \\    
(6) unspeech-128-tedx  &  30.1 & 28.7  &  29.5 & 28.0   \\   
\end{tabular}
\vspace{-4mm}
\end{table}

In Table \ref{tab:common_voice_adaption} we show WER on decoding utterances from the Common Voice V1 dev and test sets with TDNN-HMM acoustic models trained on TED-LIUM. Utterances from Common Voice are much harder to recognize, since a lot more noise and variability is present in the recordings and the recording have perceivably a much lower signal-to-noise ratio. Since they also contain over 2700 speakers each using an egregious range of microphones, they provide an excellent dev/test to test how robust the TDNN-HMM models are on out-of-domain data. Unsurprisingly, WERs are fairly high compared to the TED-LIUM test set with mostly clean and well pronounced speech. With Common Voice we observed that acoustic models trained with Unspeech embeddings consistently resulted in better WERs compared to the baselines, helping the model to adapt. Particularly pre-training Unspeech models on the Common Voice train data help a TDNN-HMM model trained on TED-LIUM to adapt to the style of Common Voice recordings. Embeddings from Unspeech models trained on TEDx will also perform better than the no context and i-vector baseline models. In contrast to the results in Table~\ref{tab:context_emb_tedlium}, in this decoding task, i-vectors in the acoustic model do not provide much of an improvement over the TDNN-HMM baseline model without context vectors.

\section{Conclusion}

Unspeech context embeddings contain and embed speaker characteristics, but supervised speaker embeddings like i-vectors would be better suited for tasks like speaker recognition or authentication. However, clustering utterances according to Unspeech contexts and using the cluster IDs for speaker adaptation in HMM-GMM/TDNN-HMM models is a viable alternative if no speaker information is available. While using Unspeech context embeddings as additional input features did not yield significant WER improvements compared to an i-vector baseline on TED-LIUM dev and test, we observed consistent WER reductions with out-of-domain data from the Common Voice corpus when we add Unspeech embeddings. This is a compelling use case of Unspeech context embedding for the adaptation of TDNN-HMM models. Better scores on the same/different speaker similarity task was not indicative of WER reduction -- our TEDx Unspeech models scored higher EERs, but were at the same time better context vectors in the acoustic models. We are currently also working on modifying the training objective to see if phonetic Unspeech embeddings can be trained using a similar unsupervised training procedure. Furthermore, we are releasing our source code and offer pre-trained models.\footnote{See \url{http://unspeech.net}, license: Apache 2.0} 

\vspace{1.1mm}

\begin{footnotesize}
\textbf{Acknowledgments.} We thank Michael Henretty from Mozilla for giving us access to Common Voice V1 speaker information.
\end{footnotesize}

\clearpage

\bibliographystyle{IEEEtran}

\bibliography{is18}

\end{document}